\documentclass[pre,twocolumn,showpacs,preprintnumbers,amsmath,amssymb,floatfix]{revtex4}

\usepackage{graphicx}
\usepackage{dcolumn}
\usepackage{bm}
\usepackage{enumerate}

\usepackage[normalem]{ulem}  
\usepackage[dvips]{color}

\begin{document}


\title{Experimental analysis of lateral impact on planar brittle material: spatial properties of the cracks}

\author{F.P.M. dos Santos$^{1,2}$}
\email{filipepms@gmail.com}
\author{V.C. Barbosa$^{1}$}
\email{valmar@if.ufrj.br}
\author{R. Donangelo$^{1,3}$}
\email{rdonangelo@gmail.com}
\author{S.R. Souza$^{1,4}$}
\email{srsouza@if.ufrj.br}
\affiliation{$^1$Instituto de F\'\i sica, Universidade Federal do Rio de Janeiro
Cidade Universit\'aria, \\CP 68528, 21941-972, Rio de Janeiro, Brazil}
\affiliation{$^2$ CEFET Qu\'\i mica de Nil\'opolis - Av. L\'ucio Tavares 237, 26530-060, Nil\'opolis, Brazil}
\affiliation{$^3$Instituto de F\'\i sica, Facultad de Ingenier\'\i a, \\
CP 30, 1100 Montevideo, Uruguay}
\affiliation{$^4$Instituto de F\'\i sica, Universidade Federal do Rio Grande do Sul,\\
Av. Bento Gon\c calves 9500, CP 15051, 91501-970, Porto Alegre, Brazil}

\date{\today}

\begin{abstract}
The breakup of glass and alumina plates due to planar impacts on one of their lateral sides is studied.
Particular attention is given to investigating the spatial location of the cracks within the plates.
Analysis based on a phenomenological model suggests that bifurcations along the cracks' paths are more likely to take place closer
to the impact region than far away from it, i. e., the bifurcation probability seems to lower as the perpendicular distance from the impacted lateral increases.
It is also found that many observables are not sensitive to the plate material used in this work, as long as the fragment multiplicities corresponding
to the fragmentation of the plates are similar.
This gives support to the universal properties of the fragmentation process reported in former experiments.
However, even under the just mentioned circumstances, some spatial observables are capable of distinguishing the material of which the plates are made and,
therefore, it suggests that this universality should be carefully investigated.
\end{abstract}

\pacs{46.50.+a, 62.20.M-}
\maketitle

\section{\label{sec:introuduction}Introduction\protect}
\label{sect:introduction}
The breakup of matter into smaller pieces, i. e. the fragmentation process, is of great interest  to the industrial community \cite{fragComminution}.
For instance, grinding, which is one of the steps in the comminution process in mining, is of  great relevance as
it appreciably impacts on the final costs.
The academic interest stems both from the association of the process with critical phenomena \cite{ODDERSH,GLASSrod, fragJapanese, PLATESprl, HERRegg} as well as from
understanding the propagation of cracks into brittle materials \cite{PMMA1996,PMMA1996prl,PMMA2005,ChandarKnauss2}.
In one of the proposed scenarios \cite{PMMA1996,PMMA1996prl,PMMA2005}, the propagation of the cracks is associated with dynamical instabilities
related to frustrated microcracks initiated from the main fracture.
This picture is very different from that \cite{ChandarKnauss2} in which flaw points ahead of the tip of the propagating crack can be randomly excited due
to the local intense stress field of the perturbation.
One of the main difficulties in this study lies on the extremely short time scale for the development of cracks \cite{failureWaves1,failureWaves2,failureWaves3}.
The very small spatial extension associated with the crack tip \cite{failureWaves1} also brings additional difficulties to the study.
Therefore, conclusive experimental analysis must be extremely detailed both in time and space. 
Despite many theoretical and experimental efforts \cite{PMMA1996,PMMA1996prl,PMMA2005,ChandarKnauss2,dynCrack2003,reviewFragmentation,dynCrackYoffe,dynCrack1,failureWaves1,failureWaves2,failureWaves3}, a clear scenario to the dynamics of
the cracks has yet to emerge.

Although of great interest from the theoretical point of view, the connection between the fragmentation process with critical phenomena has been criticized by some authors (see \cite{fragGeoSystems}, for instance) who emphasized the relevance of exploiting broad ranges of size scales as, in practical situations, lower and upper bounds to the size distribution must be present.
The existence of scale invariance clearly conflicts with this fact and requires a thorough analysis to avoid ambiguities.
Since this aspect has, to a large extent, been neglected in most analyses, further investigations are necessary in order to achieve definitive conclusions.

The theoretical model  we use in this work to interpret our fragmentation data lies on assumptions very close to those assumed in Ref.\ \cite{ChandarKnauss2}, where bifurcations occur at the tip of the propagating cracks.
We have successfully employed  it in the description of several experimental observables \cite{filipe2010} and, therefore, it is also adopted in the present work.

In Ref.\ \cite{filipe2010}, experimental results corresponding to the breakup of glass and alumina plates, due to lateral impact, were reported and discussed.
The main focus of that work was on the production of exclusive events, classified according to the impact velocity on one of the plate's laterals.
In this way, similar events were grouped according to the violence of the impact and the properties of the fragmentation process were studied.

In this work we use the same experimental setup of the former study but we now focus on the spatial location of the cracks.
With the present analysis, we aim at investigating the existence of preferred regions to the development of fractures and examine whether it is possibly related to standing waves caused by the impact.
We also found that the violence of the impact does not seem to be the best choice to single out events.
By collecting them according to the fragment multiplicity, we show that several properties of the glass and alumina plates are very similar, in contrast to what was obtained in the former work \cite{filipe2010} when the events were classified according to the impact velocity.

The remainder of the manuscript is organized as follows.
We briefly sketch our experimental apparatus in Sec.\ \ref{sec:setup}, whereas the main features of the model are recalled in Sec.\ \ref{sec:model}.
The results are discussed in Sec.\ \ref{sec:results} and the main findings are summarized in Sec.\ \ref{sec:conclusions}.

\section{\label{sec:setup}The experimental setup \protect} 
The fragmentation of alumina (0.5 mm thickness) and glass (1.0 mm thickness) square plates of 100.0 mm side have been studied using the apparatus presented in Ref.\ \cite{filipe2010}.
We refer the reader to that work for detailed information and describe here only its main features.
The plates are laid down on a flat surface and are laterally hit by a steel piston, which is accelerated through the release of compressed air into a pneumatic cylinder.
The air pressure is controlled by a solenoid valve, so that the impact velocity is suitably tuned, as shown in Ref.\ \cite{filipe2010}.
A steel block, whose side is larger than the plate's lateral, is attached to the end of the piston in order to ensure that, when aligned, the plates are homogeneously impacted on their lateral.

The fragments produced in the process are then collected and subsequently digitalized.
Individual fragments are identified by labeling pixels whose colors differ from the background.
In this way, the area of a fragment is directly determined by counting its contiguous pixels. 
Since the analysis is discussed in details in Ref.\ \cite{filipe2010}, we do not provide further details  here.

\begin{figure}[t]
\includegraphics[width=8.5cm,angle=0]{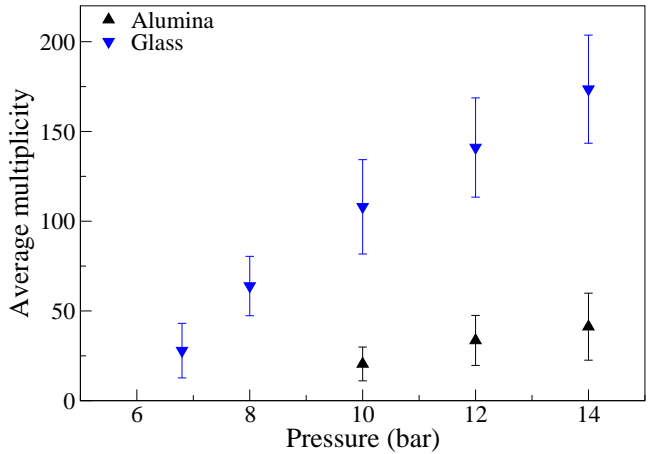}
\caption{\label{fig:presMult} (Color online) Average multiplicity as a function of the average cylinder pressure. For details see the text.}
\end{figure}

In that work,  a high resolution scanner has been employed to digitalize the fragments.
All the alumina and most of the glass data discussed in this work have been processed in this way, as they were produced in the experiments
performed in the first part of this work \cite{filipe2010}.
Extra 73 glass plates have been fragmented at low impact velocity (corresponding to the cylinder pressure equal to 6.8 bar), which gives an average fragment multiplicity equal to $M=27.9 \pm 15.2$.
As shown in Fig.\ \ref{fig:presMult}, it is only at this pressure value that there is an overlap of the fragment multiplicities for the alumina and glass data in the pressure range we study.

The new glass data has been analyzed  by photographing the fragments with a digital camera.
The perpendicular distance to the surface on which the fragments are laid is kept fixed and is adjusted to give 8.6 pixels per mm.
Since the glass plates have 1.0 mm thickness, which corresponds to 8 pixels, this resolution is sufficient for our purposes.
Furthermore, fragments whose any of its lengths has less than 10 pixels are discarded from the analyses presented below since it may correspond to the plate thickness instead of one of the fragment's side.
By discarding such fragments we leave no room for ambiguity.

\begin{figure}[bt]
\includegraphics[width=8.5cm,angle=0]{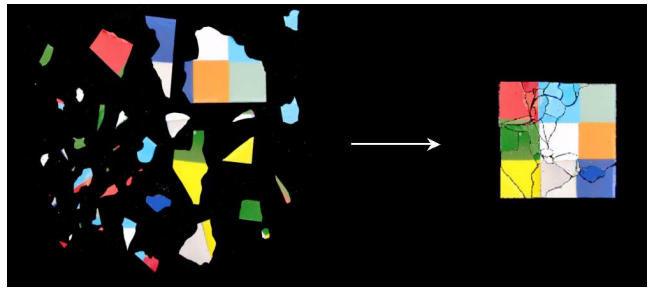}
\caption{\label{fig:reconstruct} (Color online) Reconstruction of the glass plate after the fragmentation process. 
The arrow indicates the side on which the plate was hit. For details, see the text.}
\end{figure}

The surface of these glass plates has been divided into 9 squares, which have been colored with different colors.
This helps the reconstruction process when the plate is rebuilt by arranging  the fragments (by software) one by one at their original places, as  is illustrated
in Fig.\ \ref{fig:reconstruct}.
This procedure then allows us to discuss the properties of the spatial location of the cracks.

Since all the alumina data have been taken from the former experiment \cite{filipe2010}, those plates have been reconstructed without coloring them.
Although the method employed in the analysis of the glass plates make the reconstruction much easier, it is obviously possible to do the same without colors.
Then, 66 alumina plates, which have been fragmented into $M=30.0 \pm  12.0$ pieces, have been rebuilt and will be used in the subsequent analysis.

\section{\label{sec:model}The plaque fragmentation model\protect}
The schematic model employed in this work was developed in Ref.\ \cite{filipe2010}.
As mentioned above, it is based on a scenario very close to that proposed in Ref.\ \cite{ChandarKnauss2} where secondary cracks appear at the tip of the perturbation.
Since it is detailed discussed in Ref.\ \cite{filipe2010}, we summarize its main features below:

\begin{enumerate}[(i)]
\item At the first time step, $N_c$ cracks are randomly selected at one of the plates sides and the propagation angle $\theta$ is randomly chosen between
$\pi/6 \le |\theta|\le \pi/2$, with respect to the normal to the impacted surface.
\item All cracks propagate at the same velocity at straight lines and stop only if one of the borders is met or if its course is interrupted by another crack.
\item $N_f$ flaw points, sampled from a Poisson distribution with mean value $\langle N_f\rangle$, are uniformly placed over the plates. 
They correspond to circular areas of radius $R$.
\item Secondary cracks appear with probability $P_c$ whenever the propagating crack enters the circular area of one of the flaw points.
Then its propagation direction $\theta'$ is sampled in the interval $\pi/6 \le |\theta'|\le \pi/2$ with respect to the primary crack.
\item Secondary cracks propagate following the same rules as the primary ones. 
\end{enumerate}

\noindent
As discussed in Ref.\ \cite{filipe2010}, the number of initial cracks $N_c$ and the probability of producing secondary cracks must be related and are associated with the violence of the impact.
In this work we adopt the same ad hoc functional relationship employed previously:

\begin{equation}
N_c=-10\ln(1-P_c)\;.
\label{eq:ncpc}
\end{equation}

\noindent
The parameters $\langle N_f\rangle$ and $R$, which are related to the brittleness  of the material, have been fixed in Ref.\ \cite{filipe2010} and are $\langle N_f \rangle= 10000 $,  $R=0.0005$, and the plate's area $A_0$ is set to unity.
For simplicity, this parameter set is used for both alumina and glass plates.
Thus, the only free parameter of the model is $P_c$, which is selected according to the violence of the impact.

\section{\label{sec:results}Results\protect}
\begin{figure}[t]
\includegraphics[width=8.5cm,angle=0]{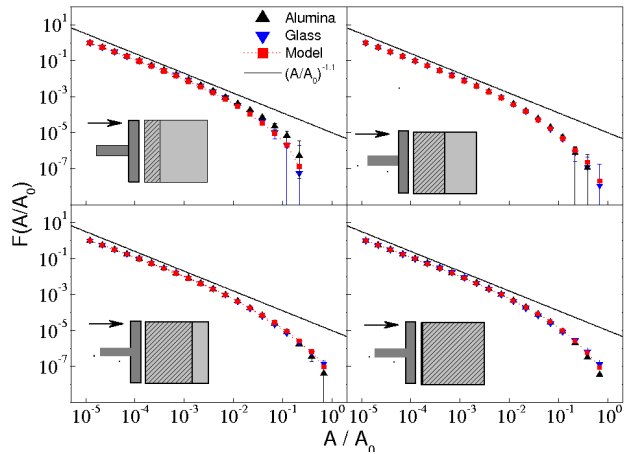}
\caption{\label{fig:sizeDist} (Color online) Size distribution for the fragmentation of alumina ($M=30.0 \pm  12.0$) and glass ($M=27.9 \pm 15.2$) plates. The hatched area, corresponding to $1/4$, $1/2$, $3/4$, and $4/4$ of the total surface, represents the region at which the centers of mass of the fragments are located. The arrows indicate the piston velocity. For details, see the text.}
\end{figure}

Fragmentation events of alumina and glass plates whose fragment multiplicities respectively correspond to  $M=30.0 \pm  12.0$ and $M=27.9 \pm 15.2$ have been grouped for analysis.
The area distribution $F(A)$ defined as \cite{ODDERSH}

\begin{equation}
F(A)=\frac{1}{A}\int_{A}^{\infty}n(A^{\prime})dA^{\prime} \;,
\label{F(m)}
\end{equation}

\noindent
where $n(A)dA$ is the number of fragments with area between $A$ and $A+dA$, is displayed in Fig.\ \ref{fig:sizeDist}.
In order to investigate whether it varies along the surface of the plates, the frames in this figure show the size distribution of fragments whose centers of mass lie inside the hatched area.
The arrows indicate the direction of the piston motion.
One observes that, except for large areas $A/A_0 \gg 0.01$, the size distributions exhibit no dependence on the position with respect to the impact region and are fairly well described by a power law $F\propto (A/A_0)^{-1.1}$ at small areas.
The sensitivity observed at large areas is obviously related to geometrical constraints.
The differences between the fragmentation of the alumina and glass plates for $A/A_0 \gg 0.01$ are due to the poor statistics at large areas and stay within the error bars.
The same remarks hold for the comparison with the model simulation, also displayed in this figure, using the model presented in Sec.\ \ref{sec:model} with $P_c=0.365$.
The interpretation of the properties of the size distribution was discussed in Ref.\ \cite{filipe2010} and the present analysis brings no new information on this context.
Since this point is not the main focus of the present work, we refer the reader to that paper for a thorough discussion.

\begin{figure}[t]
\includegraphics[width=8.5cm,angle=0]{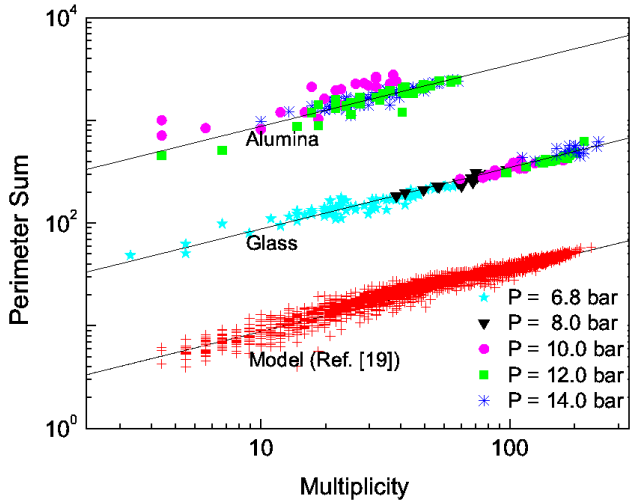}
\caption{\label{fig:perimeter} (Color online) Sum of the perimeters of all fragments for a given event as a function of the corresponding fragment multiplicity.
The model calculations have been carried out using $P_c=0.365$, 0.425, and 0.670.
The alumina data have been multiplied by $10^2$ and the glass data by $10^1$, in order to separate the sets.
The lines represent the parameterization $P_{\rm sum}/\sqrt{A_0}=2.2M^{0.6}$, also multiplied by the factors above, where necessary.
For details, see the text.}
\end{figure}

On the other hand, it is important to emphasize the independence on the plate's material observed in Fig.\ \ref{fig:sizeDist}, in agreement with the results reported in Ref. \cite{ODDERSH}.
However, in a previous work \cite{filipe2010}, we found that, for a given impact velocity, the fragmentation of alumina and glass plates does lead to distinct size distributions.
This is readily confirmed by the results shown in Fig.\  \ref{fig:presMult},  from which one clearly sees that the fragment multiplicities differ appreciably at a given impact velocity (cylinder pressure) for different plate's material.
This is a mere consequence of the fact that a given impact produces more or less damage according to the material being hit.
We find that the alumina and glass size distributions are statistically equivalent only if the events have similar multiplicities, as those selected for the present analysis.
This suggests that distinct energy amounts are needed to produce a fracture inside different materials, as is intuitively expected on physical grounds.
Indeed, as is shown in Fig.\ \ref{fig:perimeter}, the sum of the perimeters of the fragments $P_{\rm sum}$ of a given event is strongly correlated with the corresponding total fragment multiplicity $M$.
These results seem to indicate that the total fragment  multiplicity (or the energy effectively used in creation of the fractures) is the key quantity to the fragmentation process, in the pressure range studied in this work.
One should also note that there seems to be a universal function relating $P_{\rm sum}$ and $M$ as the alumina and glass data follow the same curve in a broad pressure range.
The same is true for the model results, whose points associated with different values of $P_c$ follow this curve very closely.

\begin{figure}[t]
\includegraphics[width=8.5cm,angle=0]{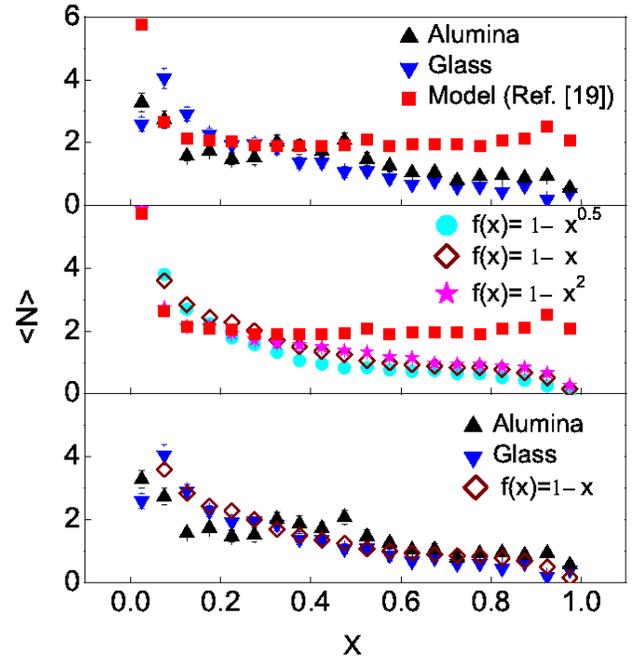}
\caption{\label{fig:n} (Color online) Number of fragments whose centers of mass lie inside the bin of length $s/s_0=0.05$ and width $s_0$, as a function of the distance $x=d/s_0$ from the impacted lateral.
For details, see the text.}
\end{figure}

As may be noted from Fig.\ \ref{fig:perimeter}, the $P_{\rm sum}$ versus $M$ curve can be fairly well approximated by $P_{\rm sum}=2.2 M^{0.6}\sqrt{A_0}$, which is represented by the lines.
In order to illustrate the extent to which this result can be explained by a simple picture, let us assume that $M-1$ fractures are created on the impacted side of a square plate of unity area and that the cracks propagate perpendicularly to this side.
In this particular model, all the fragments are rectangles of perimeter $2(1+w_i)$, where $w_i$ denotes the width of the i-th rectangle.
Since, for a  given event, the sum of $w_i$ over all fragments must add up to 1, one has $P_{\rm sum}=2(M+1)$.
Thus, considering large multiplicities, one obtains $P_{\rm sum}=2M$.
This expression differs from the empirical function found above mainly from the power law exponent.
Another simple result may be obtained by assuming that the unity area square plate is fragmented into $M$ squares.
In this case, one has $P_{\rm sum}=4M^{1/2}$, which gives an exponent much closer to the empircal value but the global factor is about two times larger than the experimental one.
The differences between these formulae and the empirical result are very likely due to the more complex geometrical shapes which are allowed in our model and experiment than in these simplified calculations.

Important insight on the underlying physics of the fragmentation properties may be gained by examining the behavior of the average number of fragments $\langle N\rangle$ whose centers of mass lie within the bins of length $s/s_0=0.05$ and width $s_0$, where $s_0=\sqrt{A_0}$.
In the top panel of Fig.\ \ref{fig:n},  $\langle N\rangle$ is plotted versus  $x=d/s_0$, where $d$ is the distance from the impacted lateral.
The model predictions are compared with the experimental alumina and glass data.
As is intuitively expected, the experimental multiplicity diminishes as one departs from the impacted border.
The model, on the other hand, predicts that $\langle N\rangle$ first falls off quickly in the neighborhood of the impacted border, and then it reaches an almost constant value all the way to the other side of the plate.
This clearly conflicts with the behavior observed experimentally.

\begin{figure}[t]
\includegraphics[width=8.5cm,angle=0]{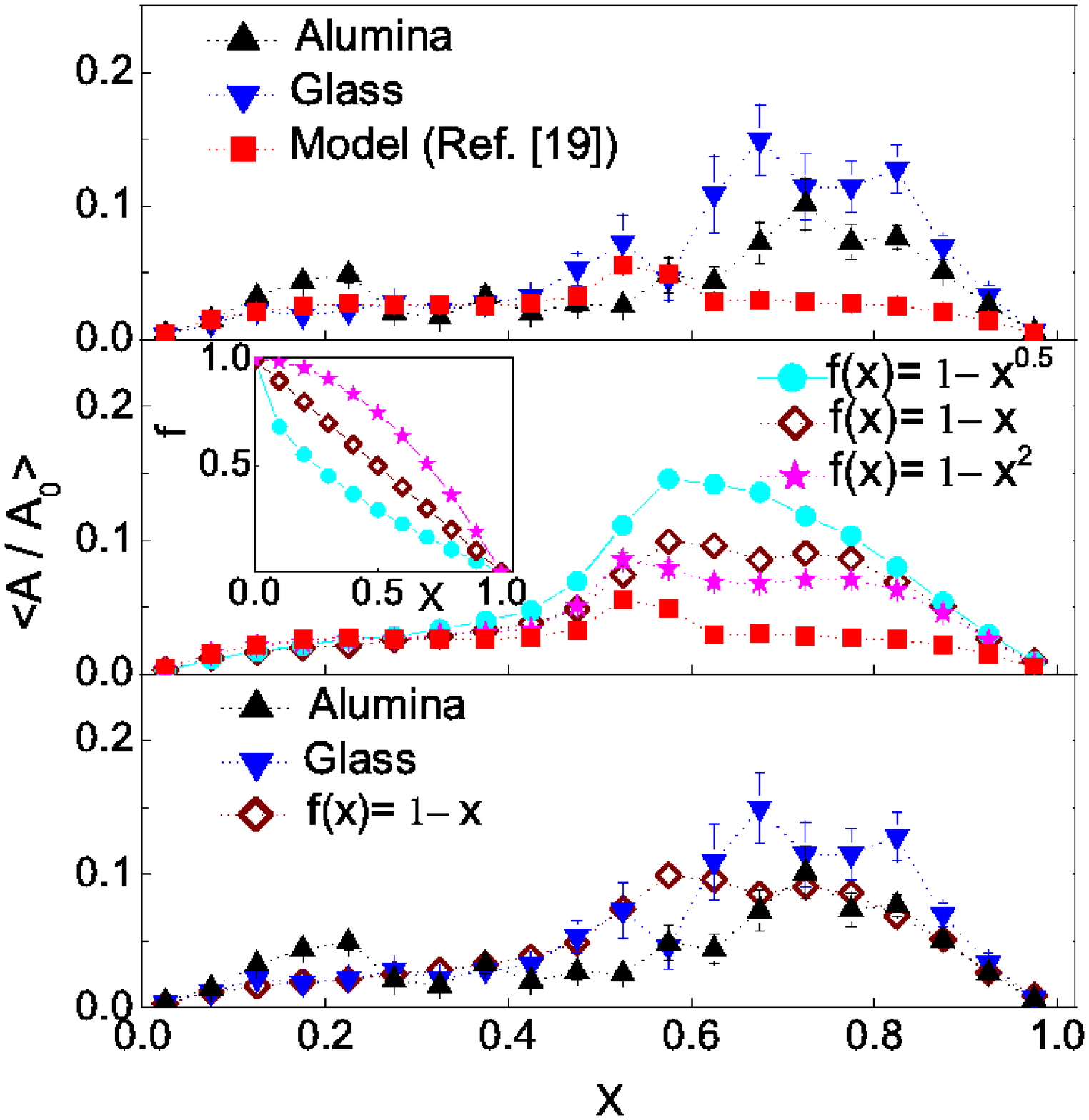}
\caption{\label{fig:aMean} (Color online) Average number of fragments whose centers of mass lie inside the bin of length $s/s_0=0.05$ and witdh $s_0$ as a function of the distance $x$ from the impacted lateral.
For details, see the text.}
\end{figure}

A possible explanation to this shortcoming is that the model does not take into account the fact that the probability of creating new cracks must drop down as one goes away from the impacted lateral since more and more energy is expended during the propagation of the crack.
In order to verify this hypothesis, we have modified our model so that the probability of creating a new crack when a flaw point is met now reads $P'_c=P_cf(x)$, where $f(x)$ is a decreasing function of $x$.
Then, $f(x)$ accounts for the fact that energy is used to disrupt matter along the crack's path.
Furthermore, as also pointed out in Refs.\ \cite{PMMA1996,PMMA1996prl,PMMA2005}, energy flows through frustrated microcracks during the propagation of the main one.
These two effects reduce the energy available to the branching process.
They are phenomenologically taken into account in our model by $f(x)$.
We have used three different functions to check the sensitivity of the results to particular choices: $f(x)=1-x^{1/2}$, $1-x$, and $1-x^2$.
The corresponding results are displayed on the middle panel of Fig.\ \ref{fig:n}. 
It should be noted that, from this point on, the model calculations in which any of these three functions $f(x)$ is employed will be explicitly labeled on the figure captions.
We  keep the label ``model" to denote the results of the standard version, which corresponds to $f(x)=1$.
The results shown in the middle panel of Fig.\ \ref{fig:n}  reveal that $N$ is not appreciably sensitive to the specific function used in the calculation, as long as it decreases monotonously as a function of $x$. 
Furthermore, one also observes that the values of $N$ obtained with $P'_c$  exhibits the desired behavior since it drops as the distance from the impacted lateral increases.

The qualitative changes brought about by these modifications also turn out to improve the agreement with the data quantitatively, as is shown at the bottom panel of Fig.\ \ref{fig:n}.
The very good agreement with the experimental results strongly suggests that the probability of creating new cracks should fall off as the perturbation front departs from the impacted region.

Despite the important information obtained above, the study of $\langle N\rangle$ does not seem to allow one to obtain precise information on $f(x)$ due to its  insensitivity to the details of the function employed in the calculation.
Therefore, we investigate the average value of the fragments' area whose centers of mass lie within a bin of length $s/s_0=0.05$ and width $s_0$.
The experimental results for both alumina and glass plates are shown at the top panel of Fig.\ \ref{fig:aMean} as a function of the
distance from the impact region.
One notices that $\langle A\rangle$ is almost flat and exhibits low values for distances up to approximately half of the plate's side $s_0$.
A bump starts to rise at this region and falls off smoothly as one approaches the other side of the plate, due to the obvious geometrical constraints.
Although large values of $\langle A\rangle$ cannot appear at small values of $x$, the explanation for the existence of a fairly broad peak in the second half of the plate is not straightforward.
For instance, the standard version of the model, in which $f(x)=1$, predicts that $\langle A\rangle$ is fairly constant all over the plate's length, as is also shown at the top panel of Fig.\ \ref{fig:aMean}.

\begin{figure}[t]
\includegraphics[width=8.5cm,angle=0]{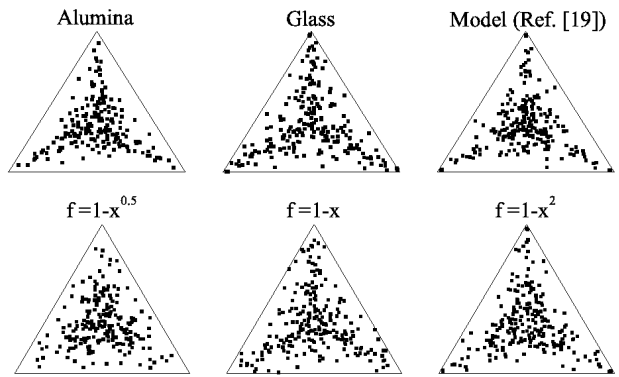}
\caption{\label{fig:Dalitz} (Color online) Dalitz plots for alumina and glass data, as well as for the different versions of the model employed in this work.
For details, see the text.}
\end{figure}

To investigate whether this bump may be explained by the model if the probability of creating new cracks drops as the distance from the impacted lateral increases, we plot, in the middle panel of Fig.\ \ref{fig:aMean},  $\langle A\rangle$ obtained in the model simulation using the same functions $f(x)$ employed above.
One notes that the bump becomes broader and higher as $f(x)$ falls faster as a function of $x$ since large fragments are more likely to be preserved if $f(x)$ drops down quickly.
The inset in this panel illustrates the differences between the distinct choices of $f(x)$ used in the calculations.
This version of the model possesses the qualitative features found in the experimental data and the comparison with the latter, shown at the bottom panel of Fig.\ \ref{fig:aMean}, reveals that $f(x)=1-x$ allows one to reproduce the experimental observations fairly well.
Due to the sensitivity of this observable shown in the middle panel of this figure, other choices for $f(x)$ could lead to a better agreement with the data.
However, the precise determination of this function is not the main aim of the present study.
 We intend to point out that important information on $f(x)$ may be obtained if one thoroughly studies $\langle A\rangle $ as a function of $x$.

A close inspection on the alumina data displayed in Fig.\ \ref{fig:aMean} reveals the presence of a small peak centered at $x\approx 0.2$.
Since the error bars are small in this region, the appearance of this bump is not likely to be due to statistical fluctuations.
Thus, there might be a mechanism responsible for it, whose interpretation is beyond the scope of the present model.
One possible explanation could be the effects associated with standing waves \cite{SPAGHETTI}, but it is difficult to draw precise conclusions on this respect from the present analysis.
However, the detailed behavior of $\langle A\rangle$ seems to be sensitive to the material being fragmented, since the glass data does not exhibit this secondary peak, and further studies are necessary to achieve definitive conclusions.

It is important to mention that we have checked that the agreement with other relevant observables, presented above and in the previous work \cite{filipe2010},  remains unchanged if one uses $P'_c=P_cf(x)$, provided $P_c$ is conviniently modified.
For brevity, to illustrate this point, we show in Fig. \ref{fig:Dalitz} the Dalitz plots \cite{filipe2010} obtained experimentally, as well as those calculated with the different versions of the model.
As discussed in Ref.\ \cite{filipe2010}, it easily reveals some qualitative features of the size distributions by focussing on the relative size of the three largest fragments within each event.
More specifically, the perpendicular distance from one of the triangles' side corresponds to the fragment's size divided by the sum of the sizes of the three largest fragments.
Thus, if the three largest fragments have almost the same size, the corresponding point is found nearly at the triangle's center.
If one of them has negligible size compared to the other two fragments, the point is located close to the middle point of one of the triangle's side.
Finally, if one fragment is much larger than the other two largest fragments, the corresponding point is found near one of the triangle's corner.
We refer the reader to Ref.\ \cite{filipe2010} where the construction of the Dalitz plot is carefully discussed and illustrated.
As is shown in Fig.\ \ref{fig:Dalitz}, the same qualitative features are found in both alumina and glass data, as well as in the different model simulations.
The qualitative agreement with the experimental results and the model simulations leads one to the conclusion that the model indeed possesses many relevant features of the fragmentation process necessary to describing the observables discussed in this work.
Indeed, except for the statistical differences, the plots furnish the same information on the fragmentation process.
They all agree on the fact tha there is an important contribution of events in which one of the three largest fragments is much larger the others.
Furthermore, one also notes that the results obtained with the different functions $f(x)$ are in very good agreement with those given by the standard version of the model.

It should also be mentioned that we found, in Ref. \cite{filipe2010}, that the largest three fragments are of approximately the same size in the fragmentation of the glass plates, in contrast with the results just reported.
Therefore, we found that there is a change of regime, where the three largest pieces of the glass plates tend to be of similar sizes at high impact velocities whereas the fragmentation process preferably produces one fragment which is much larger than the others at lower impact velocities.
The former behavior was not observed in fragmentation of the alumina plates and, if it ever takes place, it might occur at higher impact velocities.

\section{\label{sec:conclusions}Concluding Remarks\protect}
The spatial location of the cracks produced in the breakup of alumina and glass plates, due to impacts uniformly applied on one of their lateral sides is studied.
The analysis of the fragment size distributions shows that this observable is not sensitive to the distance with respect to the impact region.
Furthermore, the selection of events according to their total fragment multiplicity, rather than to the violence of the impact, shows that the fragment size distribution is the same whether alumina or glass plates are fragmented.
This is in agreement with previous findings \cite{ODDERSH} where a fairly independence on the material being fragmented has being reported for this observable but, as discussed above and also shown in Ref.\ \cite{filipe2010}, the fragmentation of alumina and glass plates impacted laterally does not exhibit such a feature, unless the analyzed events have similar multiplicities.

The average area $\langle A\rangle$ and the average number of fragments $\langle N\rangle$ as a function of the distance from the impact region have also been studied.
 Both observables are constructed considering only fragments whose centers of mass lie within a bin of length $0.05s_0$ and width $s_0$, where $s_0$ is the plate's side.
We found that, in order to reproduce the experimental findings, one needs to assume that the branching probability of cracks falls off continuously as one departs from the impact region.
The branching of cracks is a microscopic process
\cite{PMMA1996,PMMA1996prl,PMMA2005,ChandarKnauss2,dynCrack2003,reviewFragmentation,dynCrackYoffe,dynCrack1,failureWaves1,failureWaves2,failureWaves3}
 whose detailed treatment is beyond the scope of the phenomenological model used in this work.
However, the corresponding effects manifest themselves macroscopically, which allows one to examine this point.
The average fragment multiplicity $\langle N\rangle$ turned out to be fairly insensitive to the details of the function employed in the model calculation to take this fact into account in  contrast with $\langle A\rangle$, which exhibits a fairly large sensitivity to the exact functional dependence assumed.
We therefore suggest that this kind of study should be systematically carried out, for other materials and for different impact velocities/multiplicities, in order to achieve precise conclusions on this valuable piece of information regarding the propagation of cracks.

\begin{acknowledgments}
We would like to acknowledge CNPq,  FAPERJ BBP grant, CNPq-PROSUL, FAPERGS,  the joint PRONEX initiatives of CNPq/FAPERJ under
Contract No.\ 26-111.443/2010 and CNPq/FAPERGS , for partial financial support.
\end{acknowledgments}

\bibliography{article}
\end{document}